\documentclass[11pt]{article}
\usepackage{moriond,epsfig}

\def\be{\begin{equation}}
\def\ee{\end{equation}}
\def\bea{\begin{eqnarray}}
\def\eea{\end{eqnarray}}

\begin{document}
\title{Recent results from NA57 on strangeness production in p-A and 
Pb-Pb collisions at 40 and 158 $A$ GeV/c}

\author{T. Virgili, for the NA57 collaboration \\
The NA57 Collaboration:\\  
F~Antinori$^{l}$, 
P~Bacon$^{e}$, 
A~Badal{\`a}$^{g}$, 
R~Barbera$^{g}$,
A~Belogianni$^{a}$, 
A~Bhasin$^{e}$, 
I~J~Blood\-worth$^{e}$, 
M~Bombara$^{i}$, 
G~E~Bruno$^{b}$,
S~A~Bull$^{e}$,
R~Caliandro$^{b}$,
M~Campbell$^{h}$,
W~Carena$^{h}$,
N~Carrer$^{h}$,
R~F~Clarke$^{e}$,
A~Dainese$^{l}$,
A~P~de~Haas$^{s}$,
P~C~de~Rijke$^{s}$,
D~Di~Bari$^{b}$,
S~Di~Liberto$^{o}$,
R~Divi\`a$^{h}$,
D~Elia$^{b}$,
D~Evans$^{e}$,
G~A~Feofilov$^{q}$,
R~A~Fini$^{b}$,
P~Ganoti$^{a}$,
B~Ghidini$^{b}$,
G~Grella$^{p}$,
H~Helstrup$^{d}$,
K~F~Hetland$^{d}$,
A~K~Holme$^{k}$,
A~Jacholkowski$^{g}$,
G~T~Jones$^{e}$,
P~Jovanovic$^{e}$,
A~Jusko$^{e}$,
R~Kamermans$^{s}$,
J~B~Kinson$^{e}$,
K~Knudson$^{h}$,
A~A~Kolozhvari$^{q}$,
V~Kondratiev$^{q}$,
I~Kr\'alik$^{i}$,
A~Krav\v c\'akov\'a$^{j}$,
P~Kuijer$^{s}$,
V~Lenti$^{b}$,
R~Lietava$^{e}$,
G~L\o vh\o iden$^{k}$,
V~Manzari$^{b}$,
G~Martinsk\'a$^{j}$,
M~A~Mazzoni$^{o}$,
F~Meddi$^{o}$,
A~Michalon$^{r}$,
M~Morando$^{l}$,
E~Nappi$^{b}$,
F~Navach$^{b}$,
P~I~Norman$^{e}$,
A~Palmeri$^{g}$,
G~S~Pappalardo$^{g}$,
B~Pastir\v c\'ak$^{i}$,
J~Pi\v s\'ut$^{f}$,
N~Pi\v s\'utov\'a$^{f}$,
R~J~Platt$^{e}$, 
F~Posa$^{b}$,
E~Quercigh$^{l}$,
F~Riggi$^{g}$,
D~R\"ohrich$^{c}$,
G~Romano$^{p}$,
K~\v{S}afa\v{r}\'{\i}k$^{h}$,
L~\v S\'andor$^{i}$,
E~Schillings$^{s}$,
G~Segato$^{l}$,
M~Sen\'e$^{m}$,
R~Sen\'e$^{m}$,
W~Snoeys$^{h}$,
F~Soramel$^{l}$
\footnote {Permanent address: University of Udine, Udine, Italy},
M~Spyropoulou-Stassinaki$^{a}$,
P~Staroba$^{n}$,
T~A~Toulina$^{q}$,
R~Turrisi$^{l}$,
T~S~Tveter$^{k}$,
J~Urb\'{a}n$^{j}$,
F~F~Valiev$^{q}$,
A~van~den~Brink$^{s}$,
P~van~de~Ven$^{s}$,
P~Vande~Vyvre$^{h}$,
N~van~Eijndhoven$^{s}$,
J~van~Hunen$^{h}$,
A~Vascotto$^{h}$,
T~Vik$^{k}$,
O~Villalobos~Baillie$^{e}$,
L~Vinogradov$^{q}$,
T~Virgili$^{p}$,
M~F~Votruba$^{e}$,
J~Vrl\'{a}kov\'{a}$^{j}$\ and
P~Z\'{a}vada$^{n}$.
}

\address{
$^{a}$ Physics Department, University of Athens, Athens, Greece\\
$^{b}$ Dipartimento IA di Fisica dell'Universit{\`a}
       e del Politecnico di Bari and INFN, Bari, Italy \\
$^{c}$ Fysisk Institutt, Universitetet i Bergen, Bergen, Norway\\
$^{d}$ H{\o}gskolen i Bergen, Bergen, Norway\\
$^{e}$ University of Birmingham, Birmingham, UK\\
$^{f}$ Comenius University, Bratislava, Slovakia\\
$^{g}$ University of Catania and INFN, Catania, Italy\\
$^{h}$ CERN, European Laboratory for Particle Physics, Geneva, Switzerland\\
$^{i}$ Institute of Experimental Physics, Slovak Academy of Science,
       Ko\v{s}ice, Slovakia\\
$^{j}$ P.J. \v{S}af\'{a}rik University, Ko\v{s}ice, Slovakia\\
$^{k}$ Fysisk Institutt, Universitetet i Oslo, Oslo, Norway\\
$^{l}$ University of Padua and INFN, Padua, Italy\\
$^{m}$ Coll\`ege de France, Paris, France\\
$^{n}$ Institute of Physics, Prague, Czech Republic\\
$^{o}$ University ``La Sapienza'' and INFN, Rome, Italy\\
$^{p}$ Dipartimento di Scienze Fisiche ``E.R. Caianiello''
       dell'Universit{\`a} and INFN, Salerno, Italy\\
$^{q}$ State University of St. Petersburg, St. Petersburg, Russia\\
$^{r}$ IReS/ULP, Strasbourg, France\\
$^{s}$ Utrecht University and NIKHEF, Utrecht, The Netherlands
%
%
}

\maketitle\abstracts{
The production of hyperons in Pb-Pb and p-Be interaction at 40 $A$\ GeV/$c$\ 
beam momentum has been measured by the NA57 experiment. Strange particle 
enhancements at 40 $A$\ GeV/$c$\ are presented and 
compared to those measured at 158 $A$\ GeV/$c$.   
Their transverse mass spectra have been studied 
in the framework of the blast-wave model.    
The multiplicity of charged particles as a function of the energy is 
also discussed.  
}

\section{Introduction}
The experimental programme with 
heavy-ion beams at CERN SPS aims at the study of hadronic matter 
under extreme conditions of temperature, pressure and energy density.  


NA57 at the CERN SPS is a dedicated 
second-generation experiment for the study of the production of 
strange and multi-strange particles  
in Pb-Pb and p-Be collisions~\cite{NA57proposal}. 
In this paper we present results on strangeness enhancements 
at 40 $A$\ GeV/$c$ and 158 $A$\ GeV/$c$.  
A study of the  transverse mass ($m_{\tt T}=\sqrt{p_{\tt T}^2+m^2}$) 
spectra   
for  $\Lambda$, $\Xi$, $\Omega$ hyperons, their antiparticles and $K^0_s$ 
measured in Pb-Pb collisions at 158 $A$\ GeV/$c$, is also discussed.  
The multiplicity of charged particles in the central rapidity region
has been measured in Pb--Pb collisions 
at both beam momenta: 158 A GeV/{\it c} and 40 A GeV/{\it c}. 
The value of $dN_{ch}/d\eta$ at the maximum and its
behaviour as a function of centrality is here presented for the first time.

\section{Analysis and results}

The NA57 apparatus  has been described in detail elsewhere~\cite{MANZ}.
The strange particle signals are extracted 
by reconstructing the weak   
decays into final states containing only charged particles,   
using geometric and kinematic constraints, 
with a method similar to that used in the WA97 
experiment~\cite{WA97PhysLettB433}. 
For each particle species we define the fiducial  
acceptance window using a Monte Carlo simulation of the apparatus and 
excluding the border regions.  

All data are corrected for geometrical acceptance and for detector and 
reconstruction inefficiencies on an 
event-by-event basis, with the 
procedure described in reference~\cite{QM02Manzari}. 


\subsection{Multiplicity measurement}

The procedure for the measurement of the multiplicity distribution and  
the determination of the collision centrality for each class 
is described in reference~\cite{Multiplicity}.
As a measure of the collision centrality we use the number of wounded nucleons
$N_{wound}$.   
The distribution of the charged particle multiplicity  
measured in Pb-Pb interactions  
has been divided into five centrality classes (0,1,2,3,4), class 0 
being the most peripheral and class 4 being the most central.  
The fractions of the inelastic cross section for the five classes  
are given in table~\ref{tab:InvMSD}.  

The charged multiplicity  measured in the central 
unit of pseudorapidity $\eta$ is also used to determine the maximum of the 
pseudorapidity distribution 
($dN_{ch}/d\eta|_{max}$). This is the variable most frequently used to 
characterize the multiplicity of the interaction; ($dN_{ch}/d\eta|_{max}$ is 
about 2\% larger than the charged multiplicity in the central unit of $\eta$. 

\begin{figure}[tb]
\centering
\resizebox{0.8\textwidth}{!}{%
\includegraphics{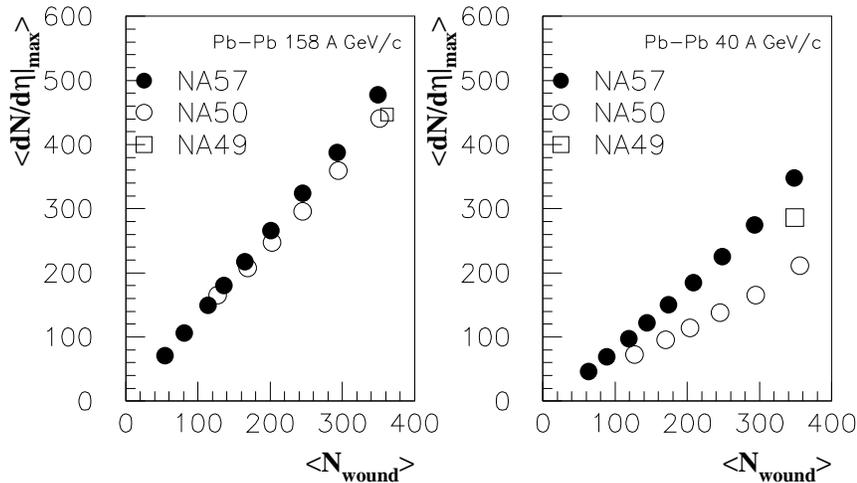}} \\
\caption{\rm $<dN/d\eta |_{max}>$ as a function of 
the average number of participants.}
\label{fig:mult1} 
\end{figure}

In fig.~\ref{fig:mult1} the values of $<dN/d\eta |_{max}>$ are reported 
as a function of $N_{wound}$ for both 40 GeV/c (right) and 158 GeV/c (left) 
beam momenta. In the same figure are reported the values  
measured by the NA50~\cite{NA50} and NA49~\cite{NA49} collaborations. 
At 158 $A$ GeV/{\it c} a reasonable agreement is observed, 
with a small discrepancy for the most central classes. 
At 40 $A$ GeV/{\it c} a strong disagreement among the three experiments 
is observed. The values of the participants for a given fraction of total 
inleastic cross-section determined by the three experiments are similar.

In proton--proton collisions, the charged multiplicity at central 
rapidity is found to scale approximately with the 
logarithm of the centre of mass energy~\cite{Eskola}.  
Assuming the same dependence one would expect: 
$dN_{ch}\eta |_{max}$(158 $A$ GeV/{\it c})/$dN_{ch}d\eta |_{max}$ (40 $A$ 
GeV/{\it c})$\simeq \ln (17.3)/\ln (8.77)$=1.31. 
The value measured in NA57 for the most central class is 1.37$\pm$0.05. 

\subsection{Transverse mass spectra in Pb-Pb at 158 $A$\ GeV/$c$}

The double-differential $(y,m_{\tt T})$\ distributions for each of the 
measured particle 
species can be parametrized using the expression  
\begin{equation}
\label{eq:expo}
\frac{d^2N}{m_{\tt T}\,dm_{\tt T} dy}=f(y) \hspace{1mm} \exp\left(-\frac{m_{\tt T}}{T_{app}}\right).
\end{equation}
Assuming the rapidity distribution to be flat within our acceptance region
($f(y)={\rm const}$),
the inverse slope parameters $T_{app}$\ (``apparent temperature'')  
have been extracted by means of maximum likelihood fits of  
equation 1 to the data. 
The $1/m_{\tt T} \, dN/dm_{\tt T} $\ distributions 
are well described by exponential functions~\cite{BlastPaper}.   

The inverse slope parameters $T_{app}$\ 
are given  in table~\ref{tab:InvMSD}  as a function of centrality, which 
is expressed for Pb-Pb intractions in terms of \% of inelastic cross section.  
\begin{table}[h]
\caption{Inverse slopes (MeV) of the  
         $m_{\tt T}$\ distributions for the five Pb-Pb centrality 
	 classes ($0$,$4$), and for p-Be and p-Pb interactions. 
	 Only statistical errors are shown.Between parenthesis the 
fractions of inelastic cross section are reported for each class.}   
\label{tab:InvMSD}
\begin{center}
\footnotesize{
\begin{tabular}{|c|c|c|cc|c|cc|} 
\hline
      &  p-Be   &   p-Pb    &    0 ($56-42$\%)     &    1 ($42-25$\%)     &    2  ($25-12$\%)    &    3  ($12-5$\%)  &    4 ($5-0$\%)  \\ \hline
$K^0_s$ &$197\pm4$& $217\pm6$ & $239\pm15$ & $239\pm8$ & $233\pm7$ & $244\pm8$ & $234\pm9$ \\
$\Lambda$  &$180\pm2$& $196\pm6$ & $237\pm19$ & $274\pm13$ & $282\pm12$ & $315\pm14$ & $305\pm15$ \\
$\bar\Lambda$ &$157\pm2$& $183\pm11$ & $277\pm19$ & $264\pm11$ & $283\pm10$ & $313\pm14$ & $295\pm14$ \\
$\Xi$ &$202\pm13$&$235\pm14$ & $290\pm20$ & $290\pm11$ & $295\pm9 $ & $304\pm11$ & $299\pm12$ \\
$\bar\Xi$&$182\pm17$&$224\pm21$ & $232\pm29$ & $311\pm23$ & $294\pm18$ & $346\pm28$ & $356\pm31$ \\
$\Omega$ + $\bar\Omega$ 
& $169\pm40$& $334\pm99$ & \multicolumn{2}{c|}{$274\pm34$} & $274\pm28$ &
        \multicolumn{2}{|c|}{$268\pm23$} \\
\hline
\end{tabular}
}
\end{center}
\end{table}

An increase of 
$T_{app}$\ 
with  
centrality is observed in Pb-Pb  
for $\Lambda$, $\Xi$  and possibly also for $\bar\Lambda$. 
Inverse slopes for p-Be and p-Pb collisions~\cite{Slope-p}  
are also given in table~\ref{tab:InvMSD}. In central  
and semi-central Pb-Pb collisions (i.e. classes 1 to 4) the baryon and 
antibaryon  $m_{\tt T}$ distributions have similar slopes.
This suggests that strange baryons and antibaryons are produced by a 
similar  mechanism.   

Within the blaste-wave model~\cite{BlastRef} the apparent temperature 
can be interpreted as due to the  
thermal motion coupled with a collective transverse flow  
of the fireball. The model predicts a double differential cross-section    
of the form:  
\begin{equation}
\frac{d^2N_j}{m_{\tt T} dm_{\tt T} dy} 
    = \mathcal{A}_j  \int_0^{R_G}{ 
     m_{\tt T} K_1\left( \frac{m_{\tt T} \cosh \rho}{T} \right)
         I_0\left( \frac{p_{\tt T} \sinh \rho}{T} \right) r \, dr}
\label{eq:Blast}
\end{equation}
where $\rho(r)=\tanh^{-1} \beta_{\perp}(r)$\ is a transverse boost,   
$K_1$\ and $I_0$\ are  modified Bessel functions, $R_G$\ is the 
transverse geometric radius of the source at freeze-out 
and $\mathcal{A}_j$\ is a normalization constant.  
The transverse velocity field $\beta_{\perp}(r)$\ has been parametrized 
according to a power law: 
\begin{equation}
\beta_{\perp}(r) = \beta_S \left[ \frac{r}{R_G} \right]^{n}  
  \quad \quad \quad r \le R_G
\label{eq:profile}
\end{equation}  
With  
this type of profile the numerical value of $R_G$\ does not 
influence the shape of the spectra but just the absolute  normalization 
(i.e. the $\mathcal{A}_j$\ constant). 
The parameters which can be extracted from a fit of equation~\ref{eq:Blast} to 
the experimental spectra are thus the thermal freeze-out 
temperature $T$\ and the 
{\em surface} 
transverse flow velocity $\beta_S$. 
Assuming a uniform particle density, the 
latter can be replaced by the {\em average} transverse flow 
velocity,  
$
<\beta_{\perp}> = \frac{2}{2+n}  \beta_S
$~\cite{BlastPaper}.   
The use of the three profiles $n=0$, $n=1/2$\ and $n=1$\ results in     
similar values of the freeze-out temperatures and of the  average transverse 
flow velocities, with good values of $\chi^2/ndf$. 
The quadratic profile is disfavoured by our data~\cite{BlastPaper}.  

The global fit of equation~\ref{eq:Blast} with $n=1$\ 
to the spectra of all the measured strange particle  
describes the data  
with $\chi^2/ndf=37.2/48$, yielding the following values for the two 
parameters  $T$\ and $ <\beta_\perp>$ for the most central class:  
$ T = 118 \pm 13  {\rm MeV} \, , \quad 
 <\beta_\perp>=0.45 \pm 0.02$.
%
The $T$\ and $<\beta_\perp>$\ parameters are statistically 
anti-correlated. The systematic errors on $T$\ 
and $<\beta_{\perp}>$\  are correlated; 
they are estimated to be $10\%$\ and $3\%$, respectively.  
%

\subsection{Strangeness enhancement}

\begin{figure}[tb]
\centering
\resizebox{0.70\textwidth}{!}{%
\includegraphics{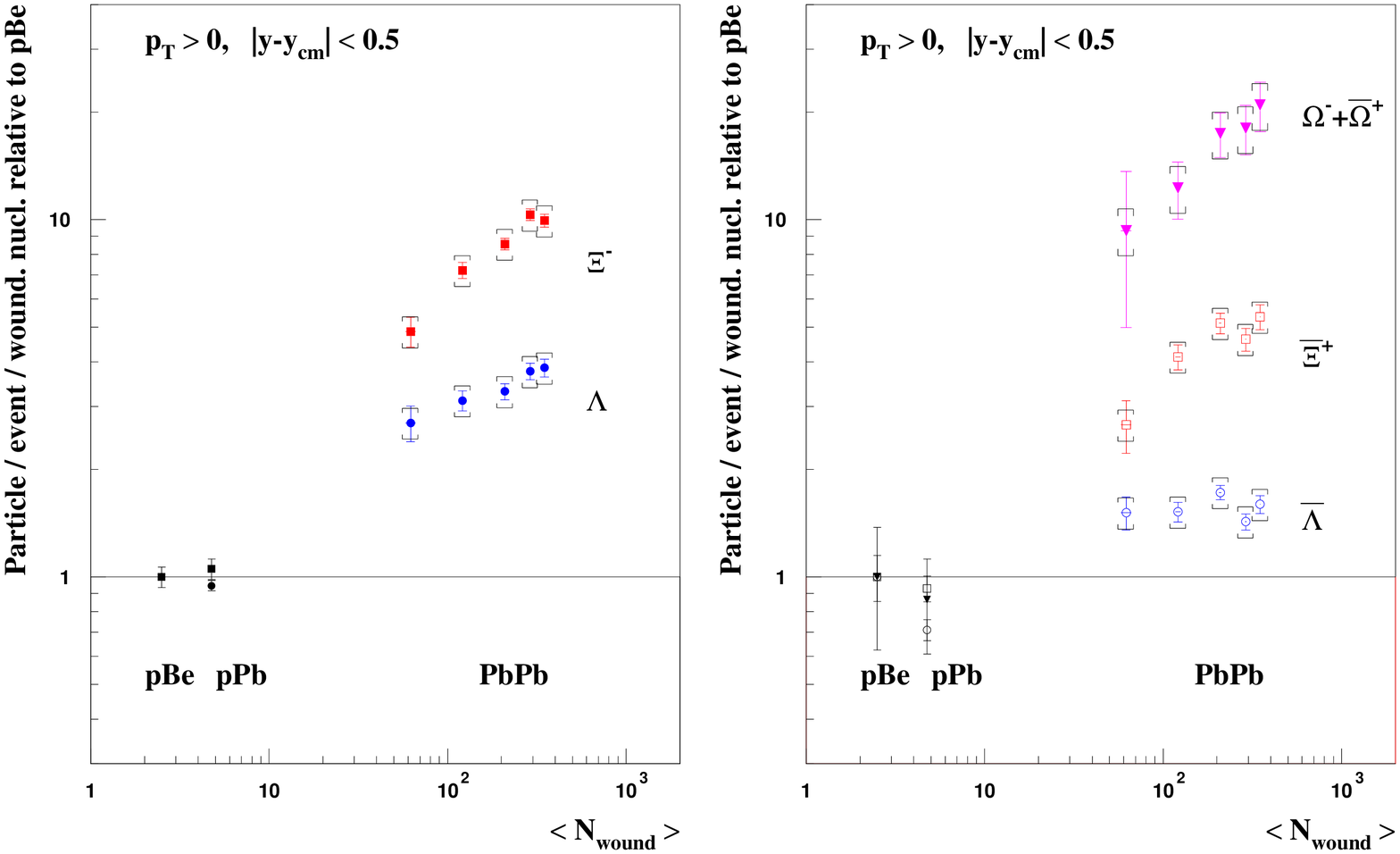}} \\
\caption{\rm 
              Hyperon enhancements $E$\  
	     as a function of the number of wounded
             nucleons at 158 $A$\ GeV/$c$. The symbol
             $_{\sqcup}^{\sqcap}$\  shows the systematic error.}
\label{fig:HypEnh1} 
\end{figure}

\begin{figure}[tb]
\centering
\resizebox{0.70\textwidth}{!}{%
\includegraphics{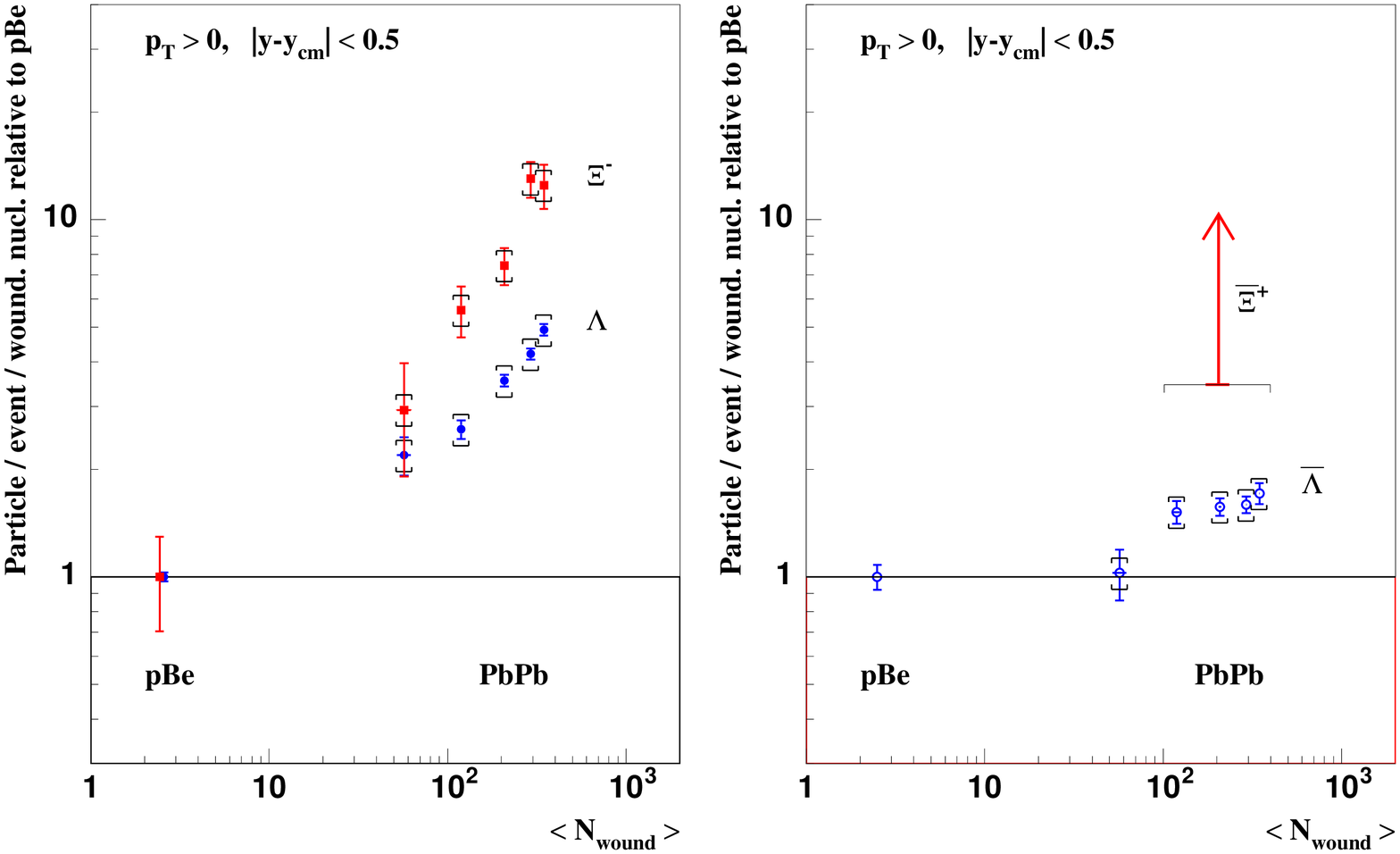}}
\caption{\rm 
              Hyperon enhancements $E$\  
	     as a function of the number of wounded
             nucleons at 40 $A$\ GeV/$c$. The symbol
             $_{\sqcup}^{\sqcap}$\  shows the systematic error.}
\label{fig:HypEnh2} 
\end{figure}

By using equation~\ref{eq:expo} we can extrapolate   
the yield measured in the selected acceptance window to a common phase space 
window covering full $p_{\tt T}$\ and one unit of rapidity centered 
at midrapidity:  
\begin{equation}
Y=\int_{m}^{\infty} {\rm d}m_{\tt T} \int_{y_{cm}-0.5}^{y_{cm}+0.5} {\rm d}y
  \frac{{\rm d}^2N}{{\rm d}m_{\tt T} {\rm d}y}.
\label{eq:yield}
\end{equation}
\noindent
The 
enhancement $E$\ 
is defined as  
\begin{equation}
E={\left(  \frac{Y}{<N_{wound}>}  \right)_{Pb-Pb}} / {
   \left(  \frac{Y}{<N_{wound}>}  \right)_{p-Be}     }
\label{eq:enh2} 
\end{equation} 
In figure~\ref{fig:HypEnh1} and figure~\ref{fig:HypEnh2} we show the 
enhancements 
as a function of $N_{wound}$\  for 158  and 
 40 $A$\ GeV/$c$ respectively.   


The enhancements are 
shown separately for particles containing at least 
one valence quark in common with the nucleon (left) and for those with 
no valence quark in common with the nucleon (right).  

The 158 $A$\ GeV/$c$\ results confirm the picture which emerged from 
WA97 --- the 
enhancement 
increases with the strangeness content of the hyperon --- 
and extend the measurements to lower centrality.  For all the particles  
except for $\bar\Lambda$ we observe a significant centrality dependence of 
the enhancements, although a saturation cannot be excluded 
for the two or three most central classes.   

A significant enhancement of strangeness production  
when going from p-Be to Pb-Pb  
is observed also in the 40 $A$\ GeV/$c$\ data. 
For the $\bar\Xi$\ particle, due to the limited  
statistics in p-Be collisions at 40 GeV/$c$, we 
could estimate only 
an upper limit to  
the production yield.  This limit for the four most central classes   
at 95\% confidence level  is indicated by 
the arrow in figure~\ref{fig:HypEnh2} (right). The enhancement pattern  
follows the same hierarchy with the strangeness content observed at 158 GeV/$c$:  
$ E(\Lambda) < E(\Xi)$\ and $E(\bar\Lambda) <  E(\bar\Xi)$.   
Comparing  the measurements at the two  
beam momenta: for the most central collisions (bins $3$\ and $4$) the 
enhancements are 
higher at 40 than at 158 GeV/$c$,    
the increase with $N_{wound}$\ is steeper at 40 than at 158 GeV/$c$.  


%
\section{Conclusions}
We have reported an enhanced production of $\Lambda$, $\bar\Lambda$, 
$\Xi$\ and $\bar\Xi$\ 
when going from p-Be to Pb-Pb collisions at 40 $A$\ GeV/$c$.  
The enhancement pattern follows the 
same hierarchy with the strangeness content as at 158 GeV/$c$: 
$ E(\Lambda) < E(\Xi)$, 
$ E(\bar\Lambda) < E(\bar\Xi)$. 
For central collisions (classes $3$\ and $4$) the enhancement is larger 
at 40 GeV/$c$.  
In Pb-Pb collisions the hyperon yields increase with $N_{wound}$\ faster at 40 than at 
158 $A$\ GeV/$c$.  

The analysis of the transverse mass spectra at 158 $A$\ GeV/$c$\ 
in the framework of the blast-wave  model 
suggests  that after a central 
collision the system expands explosively and then it 
freezes-out when the temperature is of the order  
of 120 MeV with an average transverse  flow   
velocity of about one half of the speed of light.  

Finally, the measurements of the charged particle multiplicity indicate 
that $dN_{ch}/d\eta$ at the maximum 
is close to a logarithmic scaling with the centre of mass energy.


\begin{thebibliography}{33}
%
%
\bibitem{NA57proposal} Caliandro R {\it et al.}, NA57 proposal, 1996 
{\it CERN/SPSLC 96-40, SPSLC/P300}. 
%
\bibitem{MANZ} Manzari V {\it et al.} 2001  J. Phys. G {\bf 27} 383;   \hfill\break 
T. Virgili {\it et al.} 2001 (NA57 Coll.), Nucl. Phys. A {\bf 681} 165c.
%
%
\bibitem{WA97PhysLettB433} Andersen E {\it et al.} 1998 Phys. Lett.  B {\bf 433} 209; 
\hfill\break                            
Lietava R {\it et al.} 1999 J. Phys. G  {\bf 25} 181; 
\hfill\break  
Fini R A {\it et al.} 2001 J. Phys. G  {\bf 27} 375.
\bibitem{QM02Manzari} Manzari V {\it et al.} 2003  Nucl. Phys. A {\bf 715} 140c.
\bibitem{Multiplicity} Carrer N {\it et al.} 2001 J. Phys. G {\bf 27} 391; 
\hfill\break 
 Antinori F {\it et al.} 2004 submitted to J. Phys. G.
\bibitem{NA50} M.C. Abreu {\it et al.} 2002, Phys. Lett. B {\bf 530} 33; 
\hfill\break M.C. Abreu {\it et al.} 2002, Phys. Lett. B {\bf 530} 43. 
\bibitem{NA49}  S.V. Afanasiev {\it et al.}, 2002,
 Phys.Rev. C {\bf 66} 054902; \hfill\break 
T. Anticic {\it et al.} 2004,  Phys.Rev. C {\bf 69} 024902. 
\bibitem{Eskola} K.J. Eskola, Nucl. Phys. A {\bf 698} (2002) 78.

%
\bibitem{BlastPaper} Antinori F {\it et al.} 2004 J. Phys. G {\bf 30} 823.  
\bibitem{Slope-p} Fini R A {\it et al.} 2001  Nucl. Phys. A {\bf 681} 141c.
\bibitem{BlastRef} Schnedermann E, Sollfrank J and Heinz U 1993  Phys. Rev. C
                   {\bf 48} 2462; \hfill\break 
 Schnedermann E, Sollfrank J and Heinz U
                   1994 Phys. Rev.  C {\bf 50} 1675.

%
\end{thebibliography}
\end{document}